\documentclass[]{PoS}
\newcommand{\squeezeup}{\vspace{-3.0mm}}
\newcommand{\GeV}{\ensuremath{\,\mathrm{GeV}}\xspace}

\usepackage{xspace}
\usepackage{graphicx}
\usepackage{color}
\usepackage{dcolumn}
\usepackage{bm}

\title{NLO QCD Corrections to Electroweak Higgs Boson Plus Three Jet Production at the LHC}

\ShortTitle{NLO QCD Corrections to EW $Hjjj$ Production at the LHC}

\author{Francisco Campanario \\ 
             Theory Division, IFIC, University of Valencia-CSIC \\E-46100 Paterna, Valencia, Spain\\ 
             E-mail:  \email{francisco.campanario@ific.uv.es}}
             
\author{\speaker{Terrance M. Figy} \\ 
School of Physics and Astronomy, The University of Manchester \\Manchester, M13 9PL, United Kingdom \\
E-mail: \email{Terrance.Figy@hep.manchester.ac.uk}}

\author{Simon Pl\"atzer \\
Theory Group, DESY  \\D-22607 Hamburg, Germany\\
E-mail:\email{simon.plaetzer@desy.de}}

\author{Malin Sj\"odahl \\
Department of Astronomy and Theoretical Physics, Lund University \\ SE-22362 Lund, Sweden\\
\email{malin.sjodahl@thep.lu.se}}

\abstract{The implementation of the full next-to-leading order (NLO) QCD corrections to electroweak Higgs boson plus three jet production at hadron colliders such as the LHC within the {\tt Matchbox} NLO framework of the {\tt Herwig++} event generator is discussed.  We present numerical results for integrated cross sections and kinematic distributions. }

\FullConference{11th International Symposium on Radiative Corrections (Applications of Quantum Field Theory to Phenomenology) (RADCOR 2013),\\
		22-27 September 2013\\
		Lumley Castle Hotel, Durham, UK }

\begin{document}

\section{Introduction}

In recent reports, the ATLAS and CMS Collaborations have confirmed with greater confidence the existence of a new boson with a mass in the range of $125$--$126$ GeV and a spin different from one  \cite{Aad:2012tfa,Chatrchyan:1471016}, and suggest that the new particle exhibits production and decays similar to a Standard Model (SM) Higgs boson \cite{Higgs:1964pj,Higgs:1964ia,Englert:1964et,Guralnik:1964eu}.  Further, reports from the ATLAS and CMS Collaborations indicate that current data provide evidence for a spin--$0$ Higgs boson with positive parity \cite{Aad:2013xqa,CMS-PAS-HIG-13-005} and have performed  measurements of Higgs boson production and couplings for di--boson final states \cite{Aad:2013wqa,CMS-PAS-HIG-13-016}.

The production of a Higgs boson via vector boson fusion (VBF), i.e. the $t$-channel $\mathcal{O}(\alpha_{QED}^{3})$ reaction $qq \to qq H$, is an essential channel at the LHC
for constraining Higgs boson couplings to gauge bosons and fermions.  With the current experimental data from the LHC,  the ATLAS Collaboration find $3 \sigma$  evidence \cite{Aad:2013wqa} for Higgs boson production via VBF while the CMS Collaboration find $1.3\sigma$ evidence \cite{CMS-PAS-HIG-13-022}.  
The observation of two forward tagging jets in Higgs boson production via VBF is crucial for the reduction of backgrounds.   The additional requirement that there is no extra radiation within the rapidity gap between the forward tagging jets known as the central jet veto (CJV) proposal leads to a further suppression of QCD backgrounds~\cite{Barger:1994zq,Rainwater:1998kj,Rainwater:1999sd}.  Further, the CJV proposal has been shown to be effective in reducing contamination from gluon fusion production of Higgs boson in association of two jets (GF $Hjj$) ~\cite{DelDuca:2004wt,Forshaw:2007vb,Andersen:2008gc,Cox:2010ug,Gangal:2013nxa}. 

In order to exploit the CJV strategy  for Higgs boson coupling measurements, the reduction factor due the CJV must be accurately known.  The fraction of VBF Higgs boson events with an additional jet in the rapidity gap region, i.e, the ratio the Higgs boson plus three jet (EW $Hjjj$) production cross section to the inclusive Higgs boson plus two jet (EW $Hjj$) cross section,  between the two tagging jets provides the relevant information.  Recently,  GF $Hjjj$ production has been computed within the heavy top effective theory approximation to next-to-leading order (NLO) in perturbative QCD~\cite{Cullen:2013saa}.  The usage of the heavy top effective theory approximation for $Hjj(j)$ has been validated against $Hjj(j)$ amplitudes where the top mass dependences has been kept in Refs.~\cite{Campanario:2013mga,DelDuca:2001eu}. 

The NLO QCD corrections for $Hjjj$ via VBF were presented in Ref.~\cite{Figy:2007kv,Figy:2006vc} within the $t$-channel approximation and without the inclusion of 
pentagon and hexagon one-loop Feynman diagram topologies (Figure ~\ref{fig:ggf}, last two diagrams) and the corresponding real emission contributions, which were estimated to be at the per mille level.  Given the relevance to the determination of Higgs boson couplings, we will present results from Ref. \cite{Campanario:2013fsa}, where the full NLO QCD corrections to the $\mathcal{O}(\alpha_{s} \alpha_{EW}^{3}$) production of a Higgs boson in association of three jets for the first time had been performed.  

This proceedings is organized as follows: Section~\ref{details} provides details of our NLO calculation, Section~\ref{results} presents numerical results, and in Section~\ref{concl} we conclude.

\squeezeup
\section{Calculational Details}
\label{details}

\begin{figure}[ht]
\begin{center}
\includegraphics[scale=0.65]{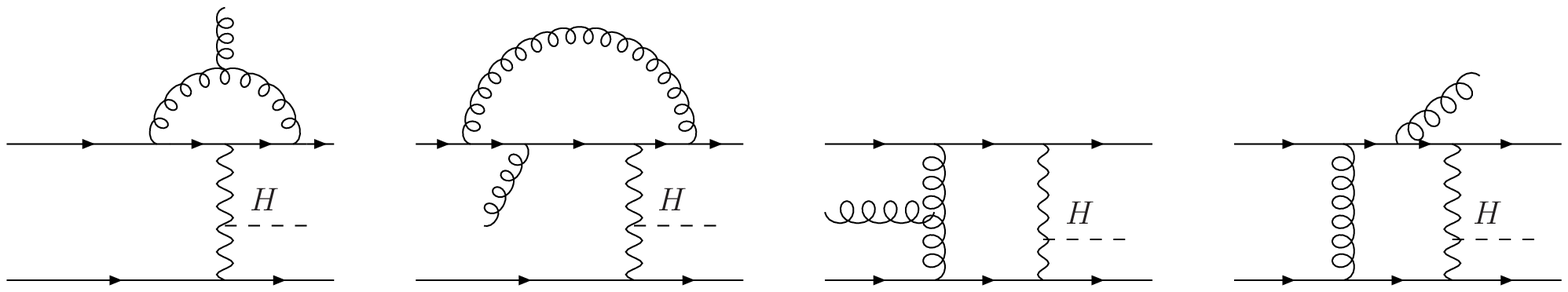}
\end{center}
\squeezeup
\caption{\label{fig:ggf}
A subset of one-loop Feynman diagram topologies for EW $Hjjj$ production.}  
\end{figure}

For the computation of the leading order (LO) $2 \to H+n$ ($n=2,3,4$) parton matrix elements, we utilized the built--in spinor helicity library of the {\tt Matchbox} module of the {\tt Herwig++} event generator~\cite{Bahr:2008pv} in order to construct the full amplitude from hadronic currents~\cite{Platzer:2011bc}.  For the computation of the interference of the virtual one--loop amplitude with the Born amplitude, we employed the helicity amplitude technique described in Ref.~\cite{Hagiwara:1988pp}. This resulted in two independent versions of the Born amplitudes which provided a valuable internal consistency check of our implementation.  
The LO $2 \to H+n$ ($n=2,3,4$) parton matrix elements were cross checked against {\tt Sherpa}~\cite{Gleisberg:2003xi,Gleisberg:2008ta} and
{\tt Hawk}~\cite{Ciccolini:2007jr,Ciccolini:2007ec}. Catani--Seymour dipole subtraction
terms \cite{Catani:1996vz} have been generated automatically by the
{\tt Matchbox} module~\cite{Platzer:2011bc}.  In order to  generate phase points more efficiently, we utilized a diagram-based multichannel phase space sampler~\cite{Platzer:2011bc}. We have used in-house routines for the one--loop virtuals, extending the techniques
developed in Ref.~\cite{Campanario:2011cs}, in order perform the reduction of the tensor integrals down to a basis of scalar one--loop integrals. The resulting amplitudes have been cross
checked against {\tt GoSam}~\cite{Cullen:2011ac}.  A representative set of one--loop Feynman diagram topologies that contribute to the virtual corrections are depicted in Figure ~\ref{fig:ggf}.

We have employed the complex mass scheme as described in Ref.~\cite{Denner:2006ic,Nowakowski:1993iu} in order to include finite width effects in gauge boson propagators.   
We use the program {\tt OneLOop}~\cite{vanHameren:2010cp} in order to compute one-loop scalar loop integrals with complex masses.  We use the
Passarino-Veltman approach \cite{Passarino:1978jh}  to reduce tensor coefficients up to four-point functions, and use the Denner-Dittmaier scheme~\cite{Denner:2005nn}, following the layout
and notation of~\cite{Campanario:2011cs} to 
numericaly evaluate the five and six point coefficients.

In order in ensure the numerical stability of our code, a test based on Ward identities has been implemented~ \cite{Campanario:2011cs}. These Ward identities are checked for each phase space point and Feynman diagram, at the expense of a small increase in computing time. If the Ward identity test fails, the amplitudes of the gauge related topology are set to zero.  The occurrence in which the Ward indenties are violated is at the per-mille level, hence, under control. 
The tensorial reduction method employed here 
has, also, been successfully applied in other scattering processes with $2 \to 4$ kinematics~\cite{Campanario:2011ud,Campanario:2013qba}. In the work presented here, the method is applied for the first time to a process which involves loop propagators with complex masses.

The color algebra associated with the computation of color correlated Born matrix elements has been performed using {\tt ColorFull}
\cite{Sjodahl:ColorFull} and cross checked using {\tt ColorMath}
\cite{Sjodahl:2012nk}. As a further check on the framework, we have implemented
the corresponding calculation of electroweak $Hjj$ production and, subsequently, 
performed cross checks against {\tt Hawk}~\cite{Ciccolini:2007jr,Ciccolini:2007ec} and VBFNLO~\cite{Arnold:2012xn}.
We have designated the implementation of the NLO corrections in perturbative QCD for electroweak Higgs boson plus two and three jet production in the {\tt Matchbox} framework as {\tt HJets++}.

\squeezeup
\section{Results}
\label{results}

The results presented here are computed for a LHC of center-of-mass energy $\sqrt{s}=14$ TeV.  
We use {\tt Herwig++}~\cite{Bahr:2008pv} to generate and analyze NLO events. We do not include parton shower and hadronization effects in our simulations. 
Hard final-state partons are recombined into jets according to the anti-$k_{T}$ algorithm~\cite{Cacciari:2008gp} 
using {\tt FastJet} \cite{Cacciari:2011ma} with $D=0.4$, $E$-scheme recombination.   
We select events with at least three jets with transverse momentum $p_{T,j} \ge
20~\rm{GeV}$ and rapidity $|y_{j}| \le 4.5$. Jets are ordered from highest to lowest in $p_{T}$.

We use the CT10~\cite{Lai:2010vv} parton distribution functions with
$\alpha_s(M_Z)= 0.118$ at NLO, and CTEQ6L1 set~\cite{Pumplin:2002vw} with $\alpha_s(M_Z)=0.130$ at LO. We use the
five-flavor scheme. We choose $m_Z=91.188 \GeV$, $m_W=80.419002 \GeV$,
$m_H=125 \GeV$ and $G_F=1.16637\times 10^{-5}\GeV^{-2}$ as electroweak
input parameters and derive the weak mixing angle $\sin \theta_{W}$
and $\alpha_{QED}$ from SM tree level relations. All
fermion masses (except the top quark) are set to zero and the CKM matrix is taken to be diagonal. 
Widths are fixed to the following values: $\Gamma_W=2.0476$ GeV and $\Gamma_Z=2.4414$ GeV.

\begin{figure}[]
\begin{center}
\includegraphics[scale=0.9]{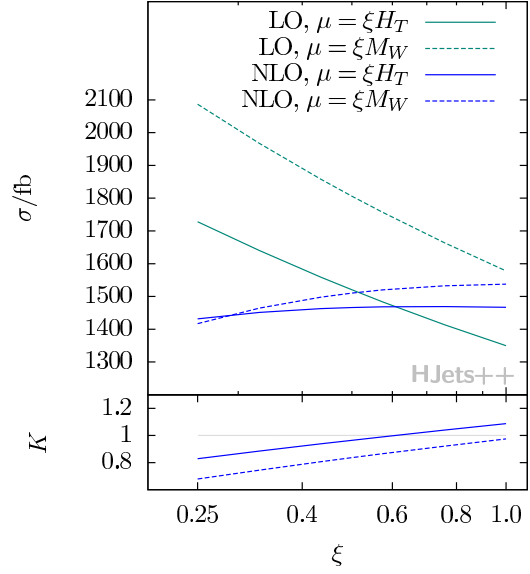}
\end{center}
\squeezeup
\caption{\label{fig:scale} The $Hjjj$ inclusive total cross section
  (in fb) at LO (cyan) and at NLO (blue) for the scale choices,
  $\mu=\xi M_{W}$ (dashed) and $\mu=\xi H_{T}$ (solid). Also, shown is
  the $K$-factor, $K=\sigma_{NLO}/\sigma_{LO}$ for $\mu=\xi M_{W}$
  (dashed) and $\mu=\xi H_{T}$ (solid).  }
\end{figure}

In Figure~\ref{fig:scale}, we show the LO and NLO total cross-sections
for inclusive cuts for different values of the factorization and
renormalization scale varied around the central scale, $\mu$ for two
scale choices, $M_W/2$, and the scalar sum of the jet transverse
momenta, $\mu_{R}=\mu_{F}= \mu= H_{T}/2$ with $H_{T}=\sum_{j}
p_{T,j}$. In general, we see a somewhat increased cross section and -
as expected - decreased scale dependence in the NLO results.  We also
note that the central values for the various scale choices are closer
to each other at NLO.  The uncertainties obtained by varying the
central value a factor two up and down are around $30\%$ ($24\%$) at LO and
$2\%$ ($9\%$) at NLO using $H_{T}/2$ ($M_W/2$) as scale choice. 
For the scale choice $\mu = H_T/2$, we obtained
$\sigma_{LO}=1520(8)^{+208}_{-171}$ fb and 
$\sigma_{NLO}=1466(17)^{+1}_{-35}$ fb.  Studying differential
distributions, we find that these generally vary less using the scalar
transverse momentum sum choice, used from now on.

\begin{figure}[ht]
\begin{tabular}{cc}
\includegraphics[scale=0.62]{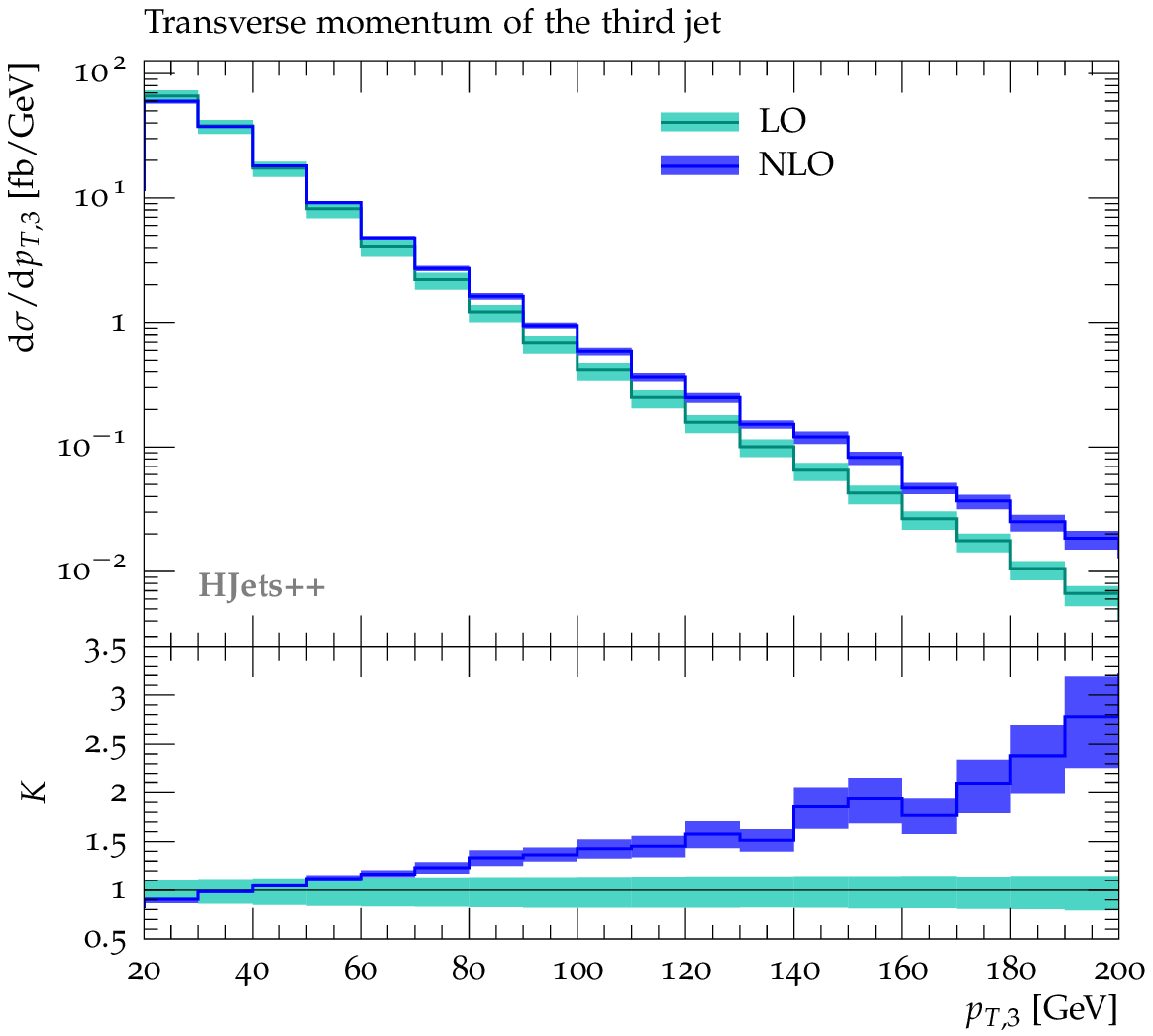} & \includegraphics[scale=0.62]{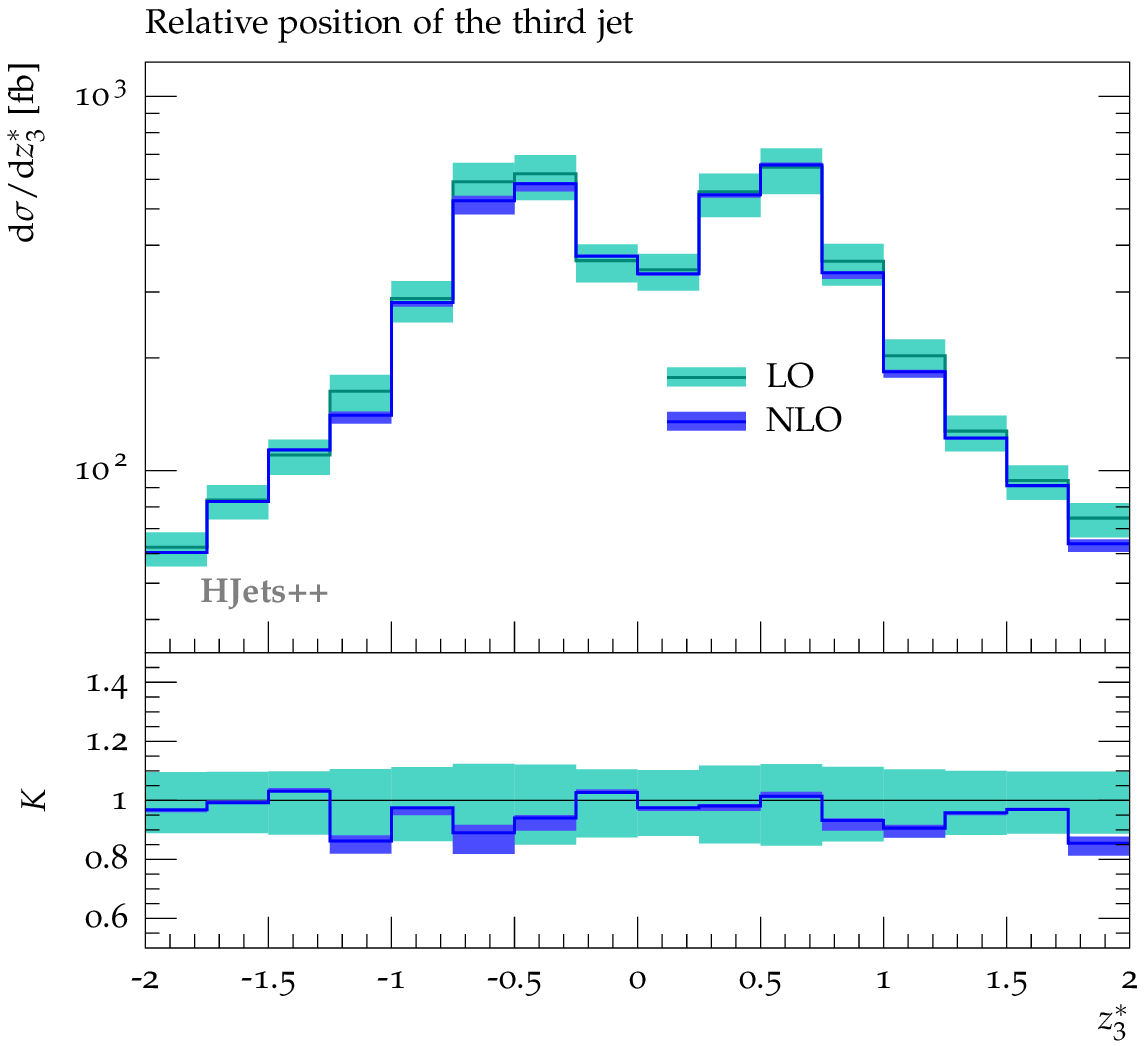} \\
\end{tabular}
\squeezeup
\caption{\label{fig:pt3} Differential cross section and $K$ factor for
  the $p_T$ of the third hardest jet (left) and the normalized centralized rapidity distribution of the third jet
  w.r.t. the tagging jets (right). Cuts are described
  in the text. The bands correspond to varying $\mu_F=\mu_R$ by
  factors 1/2 and 2 around the central value $H_{T}/2$.  }
\end{figure}

\begin{figure}[ht]
\begin{tabular}{cc}
\includegraphics[scale=0.62]{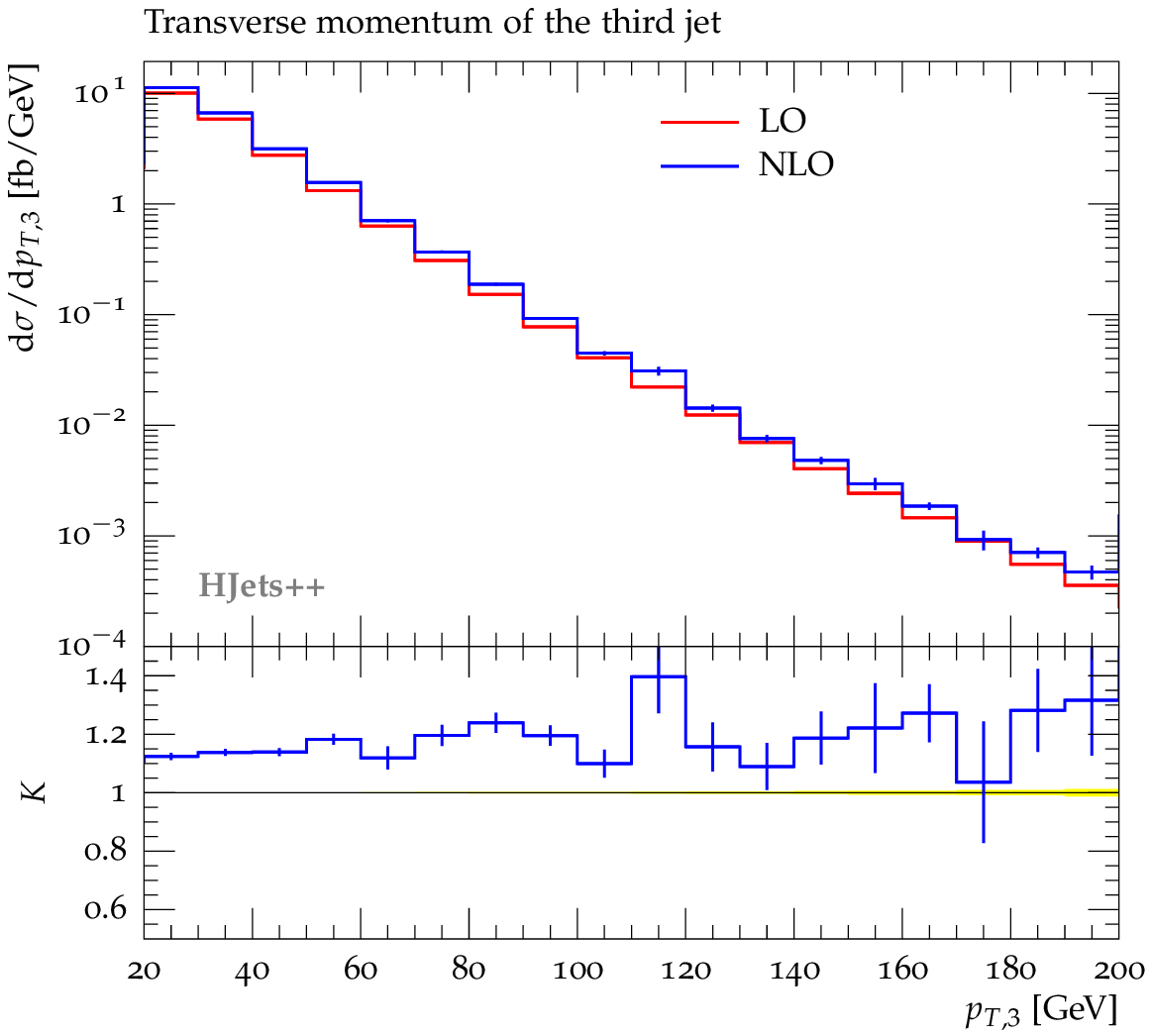} & \includegraphics[scale=0.62]{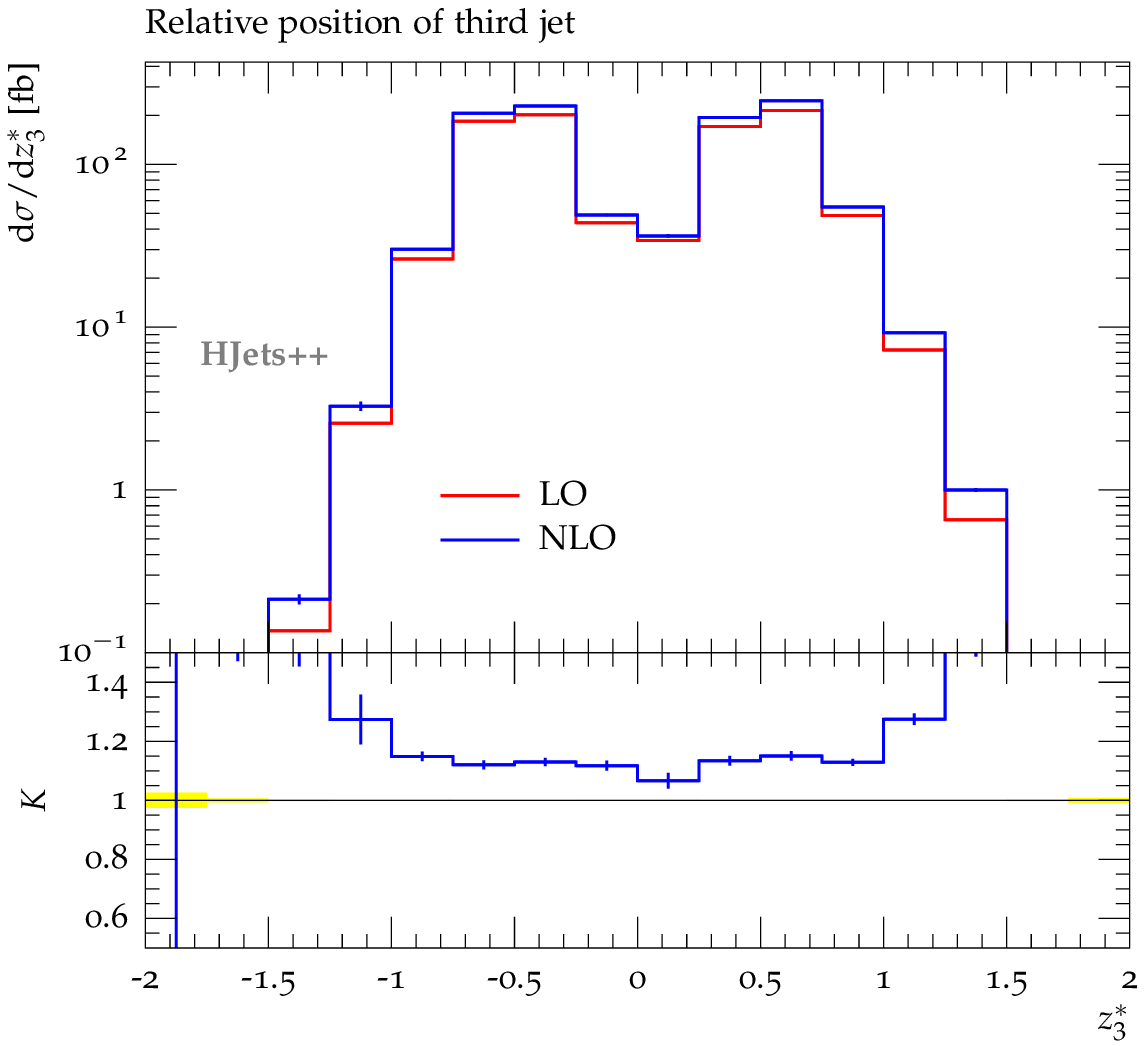}
\end{tabular}
\squeezeup
\caption{\label{fig:pt3vbf} Differential cross section and $K$ factor for
  the $p_T$ of the third hardest jet (left) and the normalized centralized rapidity distribution of the third jet
  w.r.t. the tagging jets (right) with $\mu_{R}=\mu_{F}=H_{T}$. Beyond the inclusive cuts described in the text, we include the set of VBF cuts: $m_{12} = \sqrt{(p_{1}+p_{2})^2} > 600~{\rm GeV}$ and $|\Delta y_{12}| = |y_{1} - y_{2}|> 4.0$.}
\end{figure}

On the left-hand side of Figure~\ref{fig:pt3}, the differential distribution of the third
jet, the vetoed jet for a CJV analysis, is shown. Here we find
large $K$ factors in the high energy tail of the transverse momentum
distribution. 
However, when VBF cuts \footnote{For the VBF cuts we have chosen to include the following cuts 
in addition to the inclusive cuts described in the main text : 
$m_{12} = \sqrt{(p_{1}+p_{2})^2} > 600~{\rm GeV}$ and $|\Delta y_{12}| = |y_{1} - y_{2}|> 4.0$} are included 
the $K$ factor is almost flat for the transverse momentum of the third jet (see the left-hand side of Figure~\ref{fig:pt3vbf}).
On the right-hand side of Figure~\ref{fig:pt3}, we show 
the normalized centralized rapidity distribution of the third jet w.r.t. the tagging jets,
$z^{*}_{3}=(y_3-\frac{1}{2}(y_1+y_2))/(y_1-y_2)$. This variable
beautifully displays the VBF nature present in the process.  One
clearly sees how the third jet tends to accompany one of the leading
jets appearing at $1/2$ and $-1/2$ respectively. This effect is
more pronounced when VBF cuts are applied (see Figure \ref{fig:pt3vbf}), and
should be contrasted with the gluon fusion production mechanism where
QCD radiation in the rapidity gap region between the leading two jets
will be much more common due to the $t$-channel color flow of the process
\cite{Forshaw:2007vb,Cox:2010ug,Campanario:2013mga,Cullen:2013saa}.

\squeezeup
\section{Conclusions}
\label{concl}
\squeezeup
In this proceedings, complete results at NLO QCD for electroweak Higgs boson production in association with three jets have been discussed.  The NLO corrections to the total inclusive cross section are moderate for 
inclusive cuts and the scale choice of $H_{T}/2$. However, for the scale choice of $M_{W}/2$, the NLO corrections can be more significant.  The scale uncertainty decreases from around $30\%$($24\%$) at LO down to about $2\%(${$9\%$) at NLO using the scale choice of $H_{T}/2$ ($M_{W}/2$).  We have, also, presented numerical results showing the impact of VBF selection cuts on the transverse momentum of the third jet, $p_{T,3}$, and its relative position w.r.t to the two leading jets, $z_{3}^{*}$ at NLO in perturbative QCD.

\squeezeup
\acknowledgments{We are grateful to Ken Arnold for contributions at an early stage of this project and to Mike Seymour and Jeff Forshaw for valuable discussions on the subject. F.C. is funded by a Marie Curie fellowship (PIEF-GA-2011- 298960) and partially by MINECO (FPA2011-23596) and by LHCPhenonet (PITN-GA-2010-264564). T.F. would like to thank the North American Foundation for The University of Manchester and George Rigg for their financial support. S.P. has been supported in part by the Helmholtz Alliance "Physics at the Terascale" and M.S. was supported by the Swedish Research Council, contract number 621-2010-3326.}
\squeezeup
\bibliographystyle{JHEP}
\bibliography{nlo3jets-PoS}

\end{document}